\PassOptionsToPackage{dvipsnames}{xcolor}
\documentclass[longbibliography,nofootinbib,aps,prx,superscriptaddress,twocolumn,10pt]{revtex4-2}

\usepackage{amsmath,amsfonts,amssymb,amsthm,dcolumn,bm,qcircuit,appendix,mathtools,thmtools,thm-restate,mathrsfs,pgfplots,soul,lipsum,graphicx,braket,enumitem,caption,subcaption,ragged2e}
\usepackage[linesnumbered,ruled,vlined]{algorithm2e}

\usepackage{xcolor}
\usepackage{hyperref}
\definecolor{DarkBlueHex}{HTML}{191970} % same as MidnightBlue

\usepackage[capitalize,nameinlink]{cleveref}
\hypersetup{
	colorlinks = true,
	linkcolor = [rgb]{0.70,0.13,0.13},
	citecolor = [rgb]{0.13,0.55,0.13},
	urlcolor = [rgb]{0.25, 0.41, 0.88}}

\newcommand{\Tr}{\operatorname{Tr}}
\usepackage{mathtools}
\usepackage{array}
\usepackage{multirow}
\usepackage{bbold}
\usepackage{ragged2e}%

\newcolumntype{L}[1]{>{\raggedright\let\newline\\\arraybackslash\hspace{0pt}}m{#1}}
\newcolumntype{C}[1]{>{\centering\let\newline\\\arraybackslash\hspace{0pt}}m{#1}}
\newcolumntype{R}[1]{>{\raggedleft\let\newline\\\arraybackslash\hspace{0pt}}m{#1}}

\newtheorem{definition}{Definition}

\newtheorem{theorem}{Theorem}

\newtheorem{remark}{Remark}

\newtheorem{proposition}{Proposition}

\numberwithin{proposition}{section}
\numberwithin{theorem}{section}
\numberwithin{lemma}{section}
\numberwithin{corollary}{section}
\numberwithin{remark}{section}
\numberwithin{definition}{section}
\numberwithin{equation}{section}

\DeclareRobustCommand{\orcidicon}{%
	\begin{tikzpicture}
	\draw[lime, fill=lime] (0,0) 
	circle [radius=0.16] 
	node[white] {{\fontfamily{qag}\selectfont \tiny ID}};
	\draw[white, fill=white] (-0.0625,0.095) 
	circle [radius=0.007];
	\end{tikzpicture}
	\hspace{-2mm}
}
\foreach \x in {A, ..., Z}{%
	\expandafter\xdef\csname orcid\x\endcsname{\noexpand\href{https://orcid.org/\csname orcidauthor\x\endcsname}{\noexpand\orcidicon}}
}
\begin{document}
\title{Bounding quantum uncommon information with quantum neural estimators}
\author{Donghwa Ji}
\email{donghwa722@gmail.com}
\affiliation{College of Liberal Studies, Seoul National University, Seoul 08826, Korea}
\affiliation{Team QST, Seoul National University, Seoul 08826, Korea}

\author{Junseo Lee}
\email{harris.junseo@gmail.com}
\thanks{(Current affiliation: Norma Inc.)}
\affiliation{Team QST, Seoul National University, Seoul 08826, Korea}

\author{Myeongjin Shin}
\affiliation{Team QST, Seoul National University, Seoul 08826, Korea}
\affiliation{School of Computing, KAIST, Daejeon 34141, Korea}

\author{IlKwon Sohn}
\affiliation{Quantum Network Research Center, Korea Institute of Science and Technology Information, Daejeon 34141, Korea}

\author{Kabgyun Jeong}
\email{kgjeong6@snu.ac.kr}
\thanks{Corresponding author}
\affiliation{Team QST, Seoul National University, Seoul 08826, Korea}
\affiliation{Research Institute of Mathematics, Seoul National University, Seoul 08826, Korea}
\affiliation{School of Computational Sciences, Korea Institute for Advanced Study, Seoul 02455, Korea}

\date{\today}

\begin{abstract}
In classical information theory, uncommon information refers to the amount of information that is not shared between two messages, and it admits an operational interpretation as the minimum communication cost required to exchange the messages. Extending this notion to the quantum setting, quantum uncommon information is defined as the amount of quantum information necessary to exchange two quantum states. While the value of uncommon information can be computed exactly in the classical case, no direct method is currently known for calculating its quantum analogue. Prior work has primarily focused on deriving upper and lower bounds for quantum uncommon information. In this work, we propose a new approach for estimating these bounds by utilizing the quantum Donsker–Varadhan representation and implementing a gradient-based optimization method. Our results suggest a pathway toward efficient approximation of quantum uncommon information using variational techniques grounded in quantum neural architectures.

\end{abstract}
\maketitle

\tableofcontents

\section{Introduction}

%%%고전 Intro 보충(완료)%%%

In classical information theory, a message can be modeled as a sequence of values independently drawn from a discrete probability source $X = \{(x, p_x)\}$. The entropy of the source, denoted by $H(X)$, quantifies the inherent randomness in $X$, and admits an operational interpretation as the minimal amount of information required to faithfully encode outcomes sampled from $X$.

More generally, in a base-$r$ encoding scheme, the entropy is defined as $H_r(X) = - \sum_x p_x \log_r p_x$, a quantity known as the \emph{Shannon entropy}~\cite{Sha1948EN}. Operationally, $H_r(X)$ corresponds to the expected number of $r$-ary digits needed to represent an outcome from $X$. In the special case $r = 2$, we adopt the shorthand $\log := \log_2$ and $H(X) := H_2(X)$.

Consider now a communication scenario in which a sender, Alice, wishes to transmit the source $X$ to a receiver, Bob, through a noisy communication channel $\Gamma$. The channel transforms the input $X$ into an output source $Y = \{(y, q_y)\}$, inducing a joint distribution $P_{XY}$ over $X$ and $Y$. This statistical dependence enables Bob to infer partial information about $X$ from his observation of $Y$, given knowledge of the channel. The amount of information shared between $X$ and $Y$ is captured by the \emph{mutual information}, defined as
\begin{equation}
    I(X:Y) = D_{\mathrm{KL}}(P_{XY} \Vert P_X \otimes P_Y),    
\end{equation}
where $D_{\mathrm{KL}}$ denotes the Kullback--Leibler divergence~\cite{KL1997}, and $P_X \otimes P_Y$ is the product of the marginals of $X$ and $Y$.  {This can also be expressed in terms of Shannon entropy as $I(X:Y) = H(X) + H(Y) - H(X,Y)$, which corresponds to the sum of the individual entropies minus the joint entropy.} Intuitively, mutual information quantifies the extent to which the joint distribution deviates from independence. A natural question then arises: 
\begin{center}
    \textit{``How much additional information must Bob acquire \\ in order to fully reconstruct the message $X$?''}
\end{center}

The amount of information \emph{not} shared with $Y$ is captured by the \emph{partial information} of $X$ relative to $Y$, given by $H(X) - I(X:Y)$. In classical information theory, this quantity coincides with the \emph{conditional entropy} $H(X | Y)$~\cite{Sle1971EN}.  {The operational meaning of this quantity as the minimum information required for reconstruction is formally established by the Slepian-Wolf theorem. It states that in the asymptotic limit, if Bob already possesses the correlated message Y, Alice need only send her message X at a rate of $H(X|Y)$ bits per symbol for Bob to reconstruct the message with a probability of error that approaches zero.}

\begin{figure}[t!]
\includegraphics[width=0.65\linewidth]{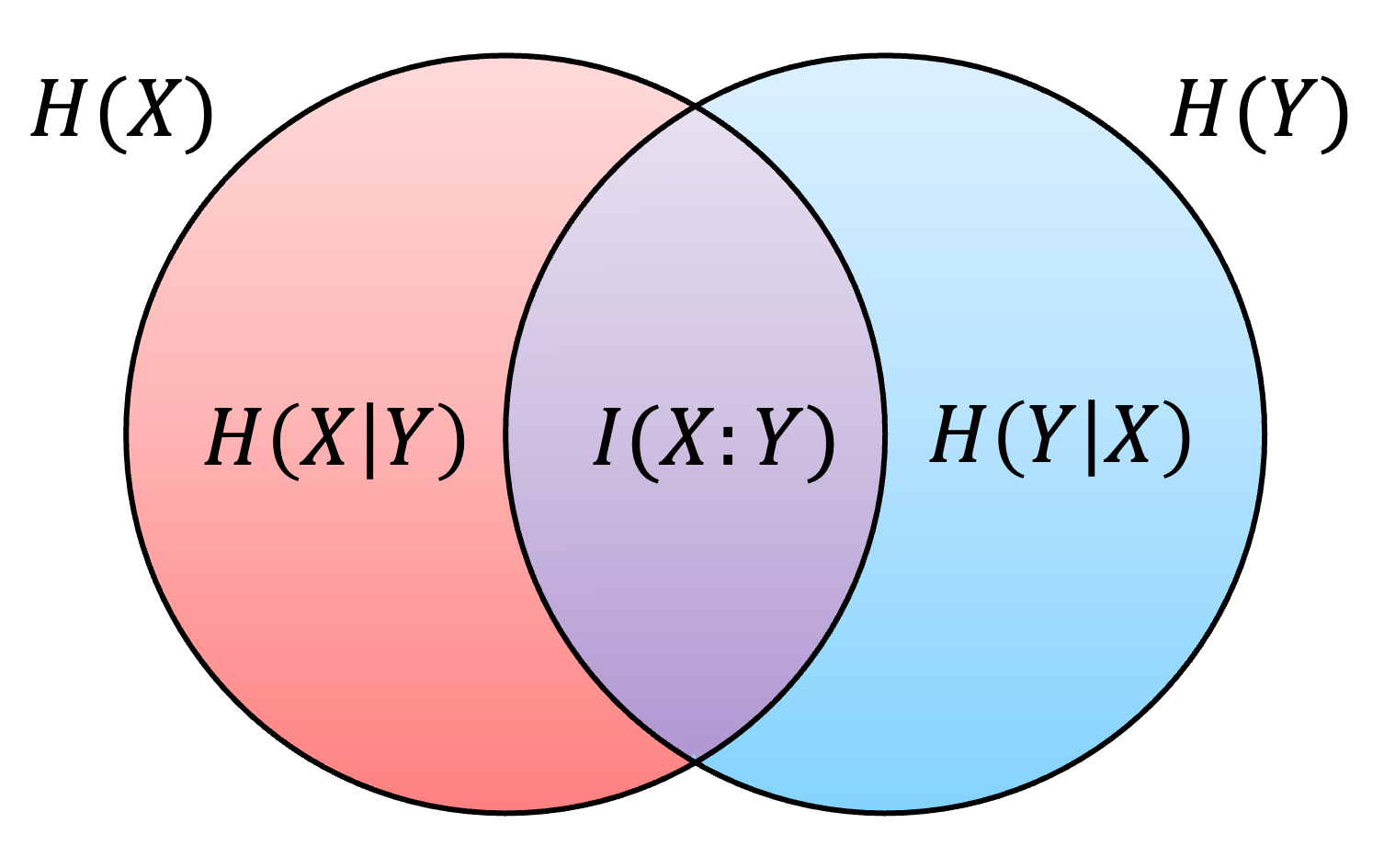}
\centering
\caption{  \textbf{Classical information quantities associated with messages $X$ and $Y$.} Each circle represents the entropy of a variable, with the overlapping region indicating the mutual information, and the non-overlapping regions corresponding to the conditional entropies.}

\label{fig_ce}
\end{figure}

Now consider a bidirectional communication setting in which Alice and Bob respectively possess correlated random variables $X$ and $Y$, and aim to exchange their messages. Due to the correlation, it suffices to exchange only the parts of the messages that are not mutually shared. The total amount of information required for this task is given by $H(X | Y) + H(Y | X)$, which is referred to as the \emph{uncommon information} between $X$ and $Y$. In classical settings, since this quantity is expressible solely in terms of conditional and mutual entropies, it is often subsumed under the broader framework of mutual information. Refer to~\cref{fig_ce} for expressions related to classical information content.

%%%고전	Entropy 그림 추가(완료)%%%

%%%양자 Intro 보충(완료)%%%

In quantum information theory, classical notions are extended to the quantum regime by replacing classical probability distributions with quantum states. The entropy of a quantum state $\rho$ is defined as $S(\rho) = - \mathrm{Tr}(\rho \log \rho)$, known as the von Neumann entropy~\cite{von1955QI}. This quantity characterizes the asymptotic number of qubits required to encode $n$ independently prepared copies of $\rho$, which is approximately $n S(\rho)$ qubits~\cite{Sch1995QI}. When a quantum system $A$ is described by the reduced state $\rho_A$, its entropy is denoted by $S(A) := S(\rho_A)$.

To extend the notion of mutual information to the quantum setting, it is crucial not to define it as the information about one system obtainable by measuring the other, since quantum measurements generally disturb the state and irreversibly alter the system. Instead, for a bipartite quantum system described by $\rho_{AB}$, where $A$ and $B$ represent subsystems held by two parties, the quantum mutual information is defined as
\begin{equation}
    I(A:B) = D(\rho_{AB} \Vert \rho_A \otimes \rho_B),    
\end{equation}
where $D(\cdot \Vert \cdot)$ denotes the Umegaki relative entropy. This quantity measures the total correlations between $A$ and $B$ by quantifying the deviation of the joint state from a product state.  {Similarly, this quantity can be written using the von Neumann entropy as $I(A:B) = S(A) + S(B) - S(AB)$.}

To formulate a quantum analogue of partial information, one may consider the quantum state merging protocol~\cite{Hor2005QSM}. Let $\ket{\psi}_{ABR}$ be a purification of the mixed state $\rho_{AB}$, where $R$ is a reference system purifying $AB$. The protocol addresses the task of transferring system $A$ from one party to the other, so that the receiving party ultimately holds the entire pure state $\ket{\psi}_{ABR}$. The quantum communication cost of this task is given by the quantum conditional entropy $S(A | B) = S(AB) - S(B)$.

Unlike its classical counterpart, this quantity can be negative. In such cases, the task can be completed without any quantum communication, and furthermore, entanglement is generated in the process. This reflects the fact that when $A$ and $B$ are highly entangled, transferring $A$ provides more than just its marginal information.  {While other positive-definite definitions of quantum relative entropy exist, they do not correspond to the operational cost of this fundamental state merging task, which provides a conceptual foundation for more elaborate protocols like state exchange.}

The concept of uncommon information also admits a quantum generalization. Extending the state merging protocol, one may consider the quantum state exchange protocol~\cite{Jon2008QUI, Lee2024QUI}, where the goal is to exchange the respective parts of the two parties so that the final global state becomes $\ket{\psi}_{BAR}$. The minimum quantum communication cost of achieving this transformation defines the quantum uncommon information.

\begin{figure}[t!]
\includegraphics[width=0.65\linewidth]{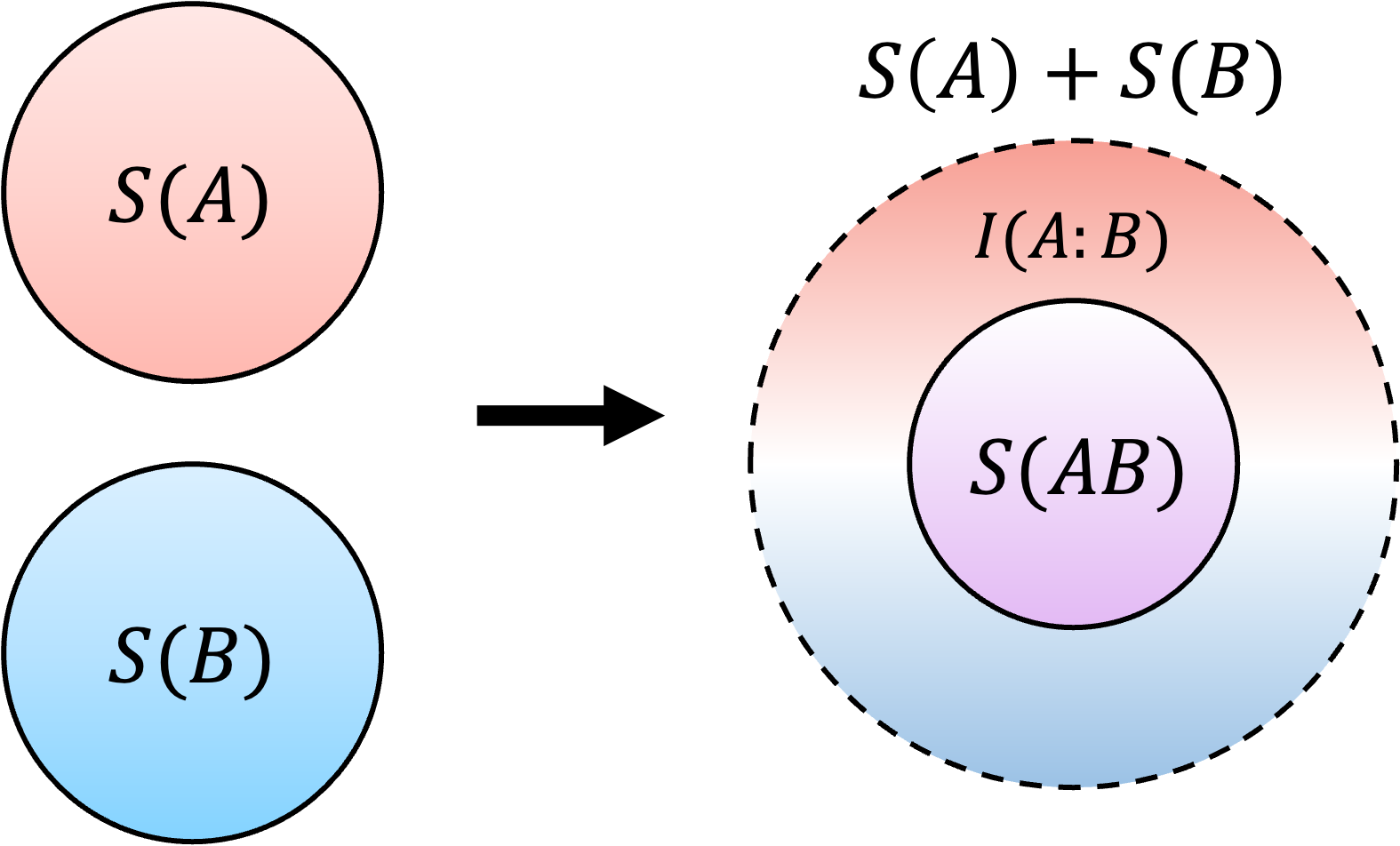}
\centering
\caption{  \textbf{Quantum information quantities associated with systems $A$ and $B$.} In the quantum setting, the total information content of the joint system $AB$ is given by the sum $S(A) + S(B)$ minus the mutual information. Unlike the classical case, the joint entropy $S(AB)$ can be smaller than either $S(A)$ or $S(B)$ due to quantum entanglement.}

\label{fig_qe}
\end{figure}

In contrast to the classical case, the quantum uncommon information cannot be expressed as $S(A | B) + S(B | A)$. Since this sum can be negative, iterating the protocol would imply the possibility of generating unbounded entanglement, which contradicts the no-cloning theorem. Therefore, no straightforward quantum analogue of the classical expression exists, and computing quantum uncommon information requires fundamentally different techniques~\cite{Hor1998QI}. Refer to~\cref{fig_qe} for expressions related to quantum information content.

%%%양자	Entropy 그림 추가(완료)%%%

The closed-form expression for the quantum uncommon information is not yet known. Although several studies have investigated specific instances~\cite{Lee2019QSE, Lee2019QSE2, Lee2021QSE}, for the general case, only upper and lower bounds have been established~\cite{Jon2008QUI, Lee2024QUI}.

 {Quantum uncommon information is not merely a theoretical construct. It serves as a key metric for quantifying communication costs in various quantum network protocols, such as distributed quantum computing~\cite{App1999DQC, App2004DQC, App2013DQC, App2016DQC}, quantum key agreement~\cite{App2017QKD, App2020QKD, App2023QKD}, and quantum secret sharing~\cite{App1999QSS, App2008QSS}.} 

In this work, we address this gap by proposing a method for predicting bounds on the quantum uncommon information of a given quantum state. Our approach leverages quantum machine learning techniques based on the quantum Donsker--Varadhan representation~\cite{Shin2024QMINE, Ziv2024QMINE}, which we employ to estimate the von Neumann entropy. These entropy estimates then enable the derivation of corresponding bounds on the quantum uncommon information.

%%%논문 구조 설명(우선 보류)%%%

\section{Quantum uncommon information}
\subsection{Quantum state exchange protocol}

The quantum uncommon information is defined operationally via the quantum state exchange protocol~\cite{Jon2008QUI, Lee2024QUI}.  {This protocol is framed within the paradigm of entanglement as a resource, where any task beyond the scope of local operations and classical communication (LOCC) must be paid for by consuming pre-shared entanglement.}

\begin{figure}[t!]
    \centering
    \includegraphics[width=0.9\linewidth]{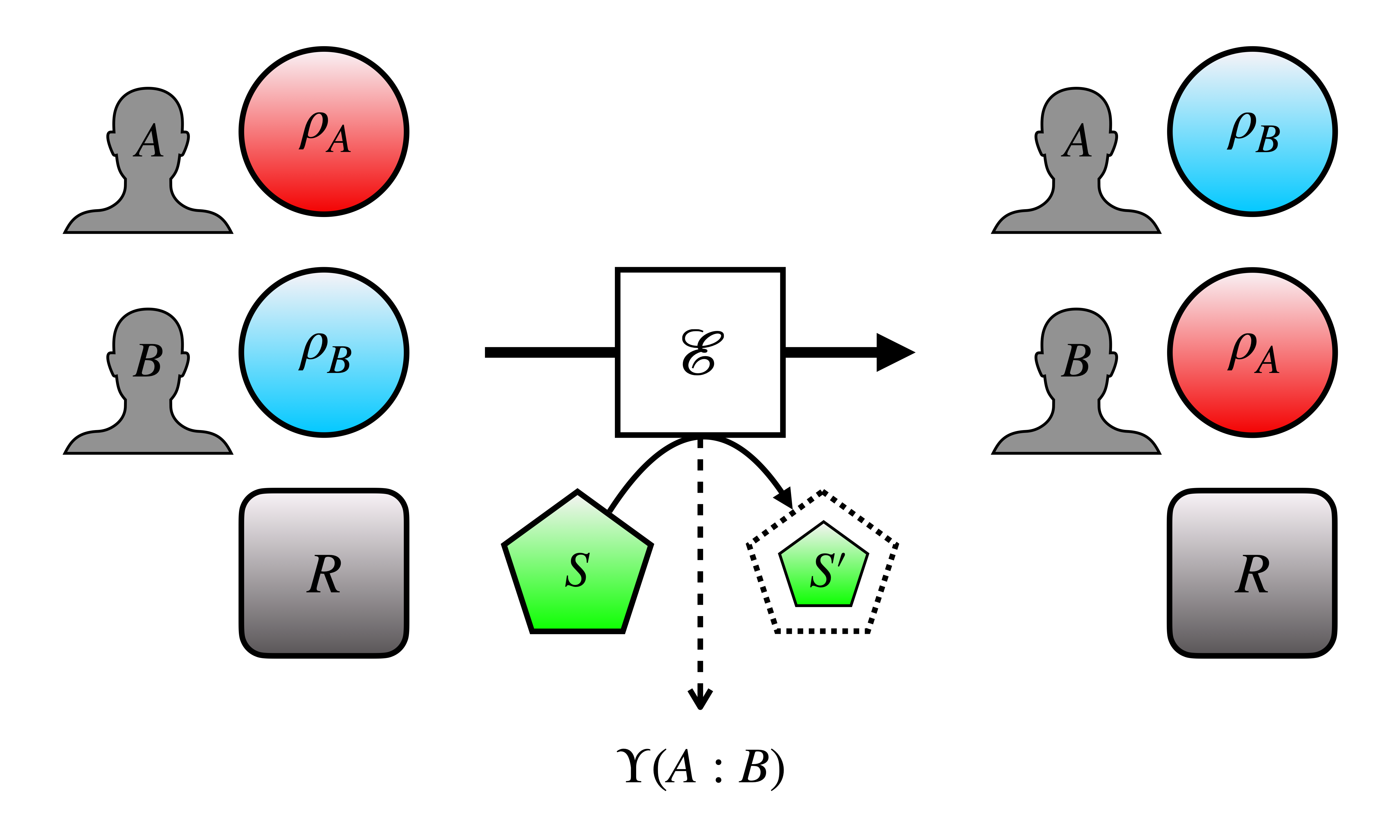}
    \caption{ 
    \textbf{Quantum state exchange protocol.} The diagram illustrates the quantum state exchange process. Alice (system $A$) holds the reduced state $\rho_A$, and Bob (system $B$) holds $\rho_B$. Their systems are exchanged via a quantum channel $\mathcal{E}$. To facilitate this process, they consume a shared maximally entangled state $S$, which is transformed into another entangled state $S'$ after the protocol. The reduction in entanglement quantifies the quantum uncommon information.}
    \label{fig_1}
\end{figure}

 {In this scenario, two parties, Alice and Bob, hold quantum systems $A$ and $B$, which are part of a larger pure state $\psi = \ket{\psi}\bra{\psi}$ on a system $ABR$, where $R$ is a reference system that purifies $AB$ such that $\Tr_{BR}(\psi) = \rho_A$ and $\Tr_{AR}(\psi) = \rho_B$. The goal is to exchange their respective systems, $A$ and $B$, using only LOCC assisted by an initial amount of shared pure entanglement. The cost of the protocol is then the net amount of entanglement, measured in units of maximally entangled states (ebits), consumed to achieve this exchange. This resource-centric view motivates the formal definition of the protocol that follows. Refer to~\cref{fig_1} for the description of the quantum state exchange protocol.}

\begin{definition}[Quantum state exchange protocol]
Let $\psi$ be a pure state on $ABR$, and let $\phi$ and $\phi'$ be maximally entangled states on auxiliary systems $S$ and $S'$, respectively. Let $\psi_{\mathrm{ex}}$ denote the target state obtained from $\psi$ by swapping systems $A$ and $B$. A {quantum state exchange protocol with error $\epsilon$} is a quantum channel
\begin{equation}
    \mathcal{E} : ABRS \rightarrow ABRS'    
\end{equation}
satisfying
\begin{equation}
    \left\| \mathcal{E} \left[ \psi \otimes \phi \right] - \psi_{\mathrm{ex}} \otimes \phi' \right\|_1 \le \epsilon,    
\end{equation}
where $\epsilon > 0$ is the permissible error.
\end{definition}

Now consider the asymptotic setting, where each party holds $n$ copies of the initial state $\psi$. If the error $\epsilon_n$ vanishes as $n \to \infty$, the protocol achieves asymptotically faithful exchange. The quantum uncommon information is then defined as the asymptotic entanglement cost of this task.

\begin{definition}[Quantum uncommon information]
Let $\psi$ be a pure state on $ABR$, and let $\psi_{\mathrm{ex}}$ be the target state with systems $A$ and $B$ exchanged. Consider a sequence of quantum channels
\begin{equation}
    \mathcal{E}_n : (ABR)^{\otimes n} \otimes S_n \rightarrow (ABR)^{\otimes n} \otimes S'_n,    
\end{equation}
where $\phi_n$ and $\phi'_n$ are maximally entangled states on $S_n$ and $S'_n$, respectively. Suppose the protocol satisfies
\begin{equation}
    \left\| \mathcal{E}_n \left[ \psi^{\otimes n} \otimes \phi_n \right] - \psi_{\mathrm{ex}}^{\otimes n} \otimes \phi'_n \right\|_1 \le \epsilon_n,    
\end{equation}
with $\epsilon_n \to 0$ as $n \to \infty$. Let $r_n$ and $r'_n$ denote the Schmidt ranks of $\phi_n$ and $\phi'_n$. Then, the quantum uncommon information $\Upsilon(A:B)$ is defined as
\begin{equation}
    \Upsilon(A:B) = \inf \left\{ \liminf_{n \to \infty} \frac{1}{n} \left( \log (r_n) - \log (r'_n) \right) \right\},    
\end{equation}
where the infimum is taken over all such sequences of protocols $\{ \mathcal{E}_n \}$ achieving vanishing error.
\end{definition}

\subsection{Bounds on quantum uncommon information}
There is no known closed-form expression for the quantum uncommon information $\Upsilon(A:B)$. However, several studies have investigated upper and lower bounds using various strategies~\cite{Jon2008QUI, Lee2024QUI}.

\subsubsection{Upper bounds}
An upper bound on $\Upsilon(A:B)$ can be obtained by constructing an explicit quantum state exchange protocol and calculating the corresponding entanglement rate. As a baseline, consider the protocol in which Alice first transmits her state $\rho_A$ to Bob, and then Bob sends his state $\rho_B$ to Alice. This approach effectively applies the quantum state merging protocol twice. The total entanglement consumption in this case is $S(A | B) + S(B) = S(AB)$, and the strategy is referred to as the \emph{merge-and-send protocol}.

This basic protocol can be improved by identifying and removing redundant components that do not require exchange. The \emph{common subspace} refers to a part of the quantum state that remains invariant under the exchange and hence need not be transmitted, reducing the total entanglement cost.

\begin{definition}[Common subspace]
Let $\ket{\psi}$ be a pure state on the tripartite system $ABR$. A subspace $C$ of both $A$ and $B$ is called a \emph{common subspace} if there exist unitary operators $U$ on $A$ and $W$ on $B$ such that
\begin{align}
\ket{\psi'} &= (U \otimes W \otimes \mathbb{1}_R) \ket{\psi} = \ket{\psi_c} + \ket{\psi_u}, \\
\ket{\psi_c} &= (\Pi_C \otimes \Pi_C \otimes \mathbb{1}_R) \ket{\psi'}, \\
\ket{\psi_u} &= (\Pi_{C^\perp} \otimes \Pi_{C^\perp} \otimes \mathbb{1}_R) \ket{\psi'}, \\
\ket{\psi_c} &= \ket{\psi_c}_{\mathrm{ex}},
\end{align}
where $\Pi_X$ denotes the projection operator onto subspace $X$, and $\ket{\psi_c}_{\mathrm{ex}}$ denotes the state $\ket{\psi_c}$ with systems $A$ and $B$ exchanged.
\end{definition}

Let $A'$ and $B'$ be ancillary systems with the same dimensions as $A$ and $B$, respectively. For fixed pure states $\ket{x} \in C^\perp$ and $\ket{y} \in C$, define the unitary operator
\begin{equation}
    U_s = \sum_{\ket{i} \in C} \ket{y} \bra{i} \otimes \ket{i} \bra{x}    
\end{equation}
acting on systems $AA'$ and $BB'$. Then the following pure state is called the \emph{stretched state}:
\begin{align}
\ket{\psi_s} &= \ket{\psi_c}_{ABR} \otimes \ket{x}_{A'} \otimes \ket{x}_{B'} \nonumber \\
&\quad + \ket{y}_A \otimes \ket{y}_B \otimes \ket{\psi_u}_{A'B'R}.
\end{align}
By converting the original state into the stretched state and applying a subspace exchange protocol that swaps only $A'$ and $B'$, the resulting entanglement cost is given by the conditional entropy $S(R \mid A)_{\psi_s}$.

\subsubsection{Lower bounds}
A lower bound on $\Upsilon(A:B)$ can be derived by observing that the sum of the initial entanglement and the entanglement used in the protocol must be at least the final entanglement.

From Alice’s perspective, the initial entanglement of system $A$ is $S(A)_{\psi}$, while after the exchange, the entanglement becomes $S(A)_{\psi_{\mathrm{ex}}} = S(B)_{\psi}$. Therefore,
\begin{equation}
    \Upsilon(A:B) \ge S(B) - S(A).    
\end{equation}
A symmetric argument from Bob’s perspective yields the lower bound
\begin{equation}
    \Upsilon(A:B) \ge |S(B) - S(A)|.    
\end{equation}

This idea can be refined by considering $n$ copies of the state and analyzing the entanglement structure across them via a decomposition protocol.

\begin{definition}[Decomposed states]
Let $\ket{\psi}$ be a pure state on $ABR$, and let $A_i$, $B_i$, and $R_i$ denote systems corresponding to Alice, Bob, and a reference, respectively. Suppose that for the $n$-fold product state $\ket{\psi}^{\otimes n}$, there exists a reversible transformation $\Lambda_n$ such that, for some error $\epsilon_n \to 0$,
\begin{align}
\Lambda_n \left[ \ket{\psi}^{\otimes n} \right] =
&\ket{\psi_1}_{A_1 R_1}^{\otimes \lfloor r_1 n \rfloor} \otimes \ket{\psi_2}_{B_1 R_2}^{\otimes \lfloor r_2 n \rfloor} \nonumber \\
&\otimes \ket{\psi_3}_{A_2 B_2}^{\otimes \lfloor r_3 n \rfloor} \otimes \ket{\psi_4}_{A_3 B_3 R_3 R_4}^{\otimes \lfloor r_4 n \rfloor},
\end{align}
where $r_i$ are non-negative rational numbers. If such a transformation exists, the four disjoint pure states $\psi_1, \dots, \psi_4$ are referred to as the \emph{decomposed states}, and the corresponding entanglement cost is given by $r_1 S(A_1)_{\psi_1} + r_2 S(B_1)_{\psi_2} + r_4 \left( S(B_3 R_3)_{\psi_4} - S(A_3 R_3)_{\psi_4} \right)$.    
\end{definition}

The best known bounds can now be summarized as follows.

\begin{proposition}[Bounds on quantum uncommon information]
Let $\Upsilon(A:B)$ denote the quantum uncommon information between quantum systems $A$ and $B$. Then the following inequalities hold:
\begin{align}
\Upsilon(A:B) &\le \inf_C \, u[C] \le S(AB), \\
\Upsilon(A:B) &\ge \sup_{\Lambda} \, l[\Lambda] \ge |S(B) - S(A)|,
\end{align}
where the upper bound is defined by $u[C] := S(R \mid A)_{\psi_s}$, and the lower bound is defined by $l[\Lambda] := r_1 S(A_1)_{\psi_1} + r_2 S(B_1)_{\psi_2} + r_4 \left( S(B_3 R_3)_{\psi_4} - S(A_3 R_3)_{\psi_4} \right)$.
\end{proposition}

\begin{remark}
    In general, it is computationally challenging to evaluate the infimum over all possible common subspaces $C$ or the supremum over all valid decompositions $\Lambda$. In this work, we focus on scenarios where the parties possess additional structural knowledge about the state, enabling efficient identification of the common subspace. For decomposed states, we restrict our attention to special cases in which such a decomposition is guaranteed. Further details are provided in~\cref{sec:est}.
\end{remark}

\section{Quantum Donsker--Varadhan representation}

The Donsker--Varadhan representation provides a lower bound on the divergence between two probability distributions~\cite{von1932DVR}. It has been widely applied to mutual information estimation by optimizing over a neural network. The quantum Donsker--Varadhan representation is the quantum analogue of this formulation, enabling entropy estimation via the Gibbs variational principle~\cite{Shin2024QMINE, Ziv2024QMINE}.

\begin{proposition}[Quantum Donsker--Varadhan representation]
Let $\rho$ be a $d$-dimensional density matrix with rank $r$, and define the function $f : \mathcal{H}^{d \times d} \to \mathbb{R}$ by
\begin{equation}
    f(T) = -\mathrm{Tr}(c \rho T) + \log\left( \mathrm{Tr}(e^{cT}) \right),    
\end{equation}
where $T$ is a Hermitian operator and $c$ is a positive constant. Then, for any $\epsilon > 0$, we have
\begin{equation}
    \left| S(\rho) - \inf_{T} f(T) \right| < \epsilon,    
\end{equation}
provided that $T$ is an $r$-rank density matrix and $c \ge 2r \log (d) - r \log (\epsilon)$.
\end{proposition}

According to this formulation, one can estimate the von Neumann entropy $S(\rho)$ by minimizing the function $f(T)$ over rank-$r$ Hermitian operators $T$. The optimal value approximates $S(\rho)$ up to arbitrary precision $\epsilon$.

\subsection{Neural estimation of quantum entropy}

Suppose $\rho$ is a density matrix of rank $r$. Let $U(\boldsymbol{\theta})$ be a parameterized unitary operator with parameters $\boldsymbol{\theta} = \{\theta_1, \dots, \theta_n\}$. We define a parameterized ansatz for the Hermitian operator $T$ as
\begin{equation}
    T(\boldsymbol{\theta}, \mathbf{t}) = \sum_{i=1}^r t_i \, U(\boldsymbol{\theta}) \ket{i} \bra{i} U^\dagger(\boldsymbol{\theta}),    
\end{equation}
where $\mathbf{t} = \{t_1, \dots, t_r\}$, $t_i \ge 0$, and $\sum_{i=1}^r t_i = 1$.

Then the function $f$ can be rewritten as
\begin{align}\label{loss}
f(\boldsymbol{\theta}, \mathbf{t}) =
&-c \sum_{i=1}^r t_i \bra{i} U^\dagger(\boldsymbol{\theta}) \rho U(\boldsymbol{\theta}) \ket{i} \nonumber \\
&+ \log \left( d - r + \sum_{i=1}^r e^{c t_i} \right).
\end{align}
The first term can be computed on a quantum computer using a parameterized quantum circuit, while the second term is evaluated classically.  {Accordingly, we used the function $f(\boldsymbol{\theta},\mathbf{t})$ from~\cref{loss} directly as the loss function. We then obtained an estimate of the von Neumann entropy by updating the parameters to minimize the value of this loss function.}

The gradient of $f$ with respect to $\mathbf{t}$ can be computed classically, and the gradient with respect to $\boldsymbol{\theta}$ can be evaluated using the parameter-shift rule~\cite{Mit2018QML}. This allows the use of hybrid quantum--classical optimization techniques, such as gradient descent, to minimize $f(\boldsymbol{\theta}, \mathbf{t})$ and thereby estimate the von Neumann entropy of $\rho$~\cite{Shin2024QMINE, Ziv2024QMINE}.

 {In our numerical simulations, we employed the Adam (Adaptive Moment Estimation) optimizer to update the variational parameters. The initial learning rate was set to 0.1, and a StepLR learning rate scheduler was also applied, which decays the learning rate by a factor of 0.9 every 10 optimization steps to enhance the stability of the training process.}

% \begin{figure}[t!]
%     \centering
%     \includegraphics[width=0.9\linewidth]{fig/fig_qml.png}
%     \caption{ 
%     \textbf{Quantum neural estimation of entropy.} The diagram illustrates the process of minimizing $f(\boldsymbol{\theta}, \mathbf{t})$ via gradient descent. The blue quantum module computes the expectation value of $\rho$ using a parameterized circuit. The orange classical module evaluates gradients and updates the parameters $\boldsymbol{\theta}$ and $\mathbf{t}$. As optimization proceeds, the function value approaches the von Neumann entropy $S(\rho)$.
%     }
%     \label{fig_qml}
% \end{figure}

It has been shown that the number of copies of $\rho$ required for this learning process scales as $O(\mathrm{poly}(r))$, where $r$ is the rank of the state~\cite{Shin2024QMINE}. This implies that the quantum Donsker--Varadhan representation is particularly effective when applied to low-rank quantum states.

\section{Estimation of bounds} \label{sec:est}

Based on the above, the quantum Donsker--Varadhan representation can be utilized to estimate the von Neumann entropy of a given density matrix. Since the bounds of quantum uncommon information are expressed in terms of von Neumann entropy, applying the same approach allows quantum Donsker--Varadhan representation to estimate the bounds of quantum uncommon information between given quantum states. Notably, as the process of expressing the bounds reduces the size of the states whose von Neumann entropy needs to be estimated, the rank of these states also decreases. Therefore, the method proves to be effective in such cases.

Consider the case of estimating the bounds of quantum uncommon information $\Upsilon(A:B)$ for quantum systems $A$ and $B$. In general, let us consider a given pure state $\psi_{ABR}$, where $R$ is a purifying system of $AB$.

\subsection{The loose bounds}

For the loose upper bound $S(AB)$ and lower bound $|S(B) - S(A)|$, the von Neumann entropy of each state can be directly estimated to obtain these values. Consequently, for loose bounds, it is relatively straightforward to approximate their values. However, for tight bounds, direct estimation becomes more challenging. 

Therefore, we impose several constraints to determine the form of the state, which will then allow us to estimate the bound value. When the tight bound can be measured, the significance of the wide bound value becomes irrelevant. However, for the purpose of comparison, we can easily estimate the bound using a state of the form $\rho_{AB}$ by applying the aforementioned method. In the following, we will explore methods for estimating tight upper and lower bounds, respectively.

\subsection{The tight bounds}

\subsubsection{Common subspaces}

Before estimating the tight upper bound, we first examine the properties of the common subspace.

Suppose the common subspace $C$ with respect to a specific basis is given by $C = \mathrm{span} \{ \ket{i_1}, \dots, \ket{i_\ell} \}$. If $\rho_A$ and $\rho_B$ satisfy the conditions for a common subspace without the need for additional unitary transformations, then when expressed in matrix form with respect to the same basis, the following must hold:
\begin{equation}
    \rho_A = \begin{pmatrix} 
                X & 0 \\ 
                0 & M_A 
            \end{pmatrix}, \quad 
    \rho_B = \begin{pmatrix} 
                X & 0 \\ 
                0 & M_B 
            \end{pmatrix},  
\end{equation}
where $X$ is an $\ell \times \ell$ matrix, and $M_A$, $M_B$ are $(n-\ell) \times (n-\ell)$ matrices. Moreover, the spectrum of $X$ must correspond to the spectrum of $\rho_A$ and $\rho_B$. Therefore, the common subspace is contained in the subspace corresponding to the components where $\rho_A$ and $\rho_B$ have the same eigenvalues.

Now suppose the density matrices are given by 
\begin{equation}
    \rho_A = \sum_{i=1}^n \alpha_i \ket{a_i}\bra{a_i}, \quad \rho_B = \sum_{i=1}^n \beta_i \ket{b_i}\bra{b_i}.
\end{equation}
We are allowed to apply arbitrary unitary operations to systems $A$ and $B$, respectively. Since a unitary operator acts as a basis change, we can align the states to a desired basis to identify the common subspace. Thus, the following holds.

\begin{proposition}[Partial spectral alignment and unitary diagonalization]\label{prop1}
Let $\rho_A$ and $\rho_B$ be density matrices of the same dimension $n$, with eigenvalues ordered as $\alpha_1 \leq \cdots \leq \alpha_n$ and $\beta_1 \leq \cdots \leq \beta_n$. Suppose there exists a permutation $\pi \in S_n$ such that for some $k$, the condition $\alpha_{\pi(i)} = \beta_{\pi(i)}$ holds for $1 \leq i \leq k$. Then there exist unitary operators $U$ and $W$ such that  
$$U \rho_A U^\dagger = \sum_{i=1}^{n} \alpha_{\pi(i)} \ket{i}\bra{i}, \quad W \rho_B W^\dagger = \sum_{i=1}^{n} \beta_{\pi(i)} \ket{i}\bra{i}.$$
\end{proposition}

Let $\{\ket{1}, \dots, \ket{k}\}$ be the basis that maximizes the overlap of eigenvalues. Then, from the above, we obtain:

\begin{theorem}[Unitary mapping of a subspace to a fixed basis segment]\label{thm3}
Let $C = \mathrm{span} \{\ket{i_1}, \dots, \ket{i_l}\}$ be a common subspace. Then there exists a unitary $U_C$ such that 
$\{ U_C\ket{i_1}, \dots, U_C\ket{i_l} \} \subset \{\ket{1}, \dots, \ket{k}\}.$
\end{theorem}

Now consider a decomposition of $\psi_{ABR}$ into subspaces $C \otimes C$ and $C^\perp \otimes C^\perp$:
$$\ket{\psi} = \sum_{i,j=1}^d \sum_{k} c_{ijk}\ket{i}\ket{j}\ket{r_k} + \sum_{i,j=d+1}^n \sum_{k} c_{ijk}\ket{i}\ket{j}\ket{r_k},$$
where $\{ \ket{r_k} \}$ is an orthonormal basis of $R$. In general, we have:

\begin{proposition}[Characterization of common subspaces via a nonzero-structure relation]\label{prop2}
Define a relation $\sim$ on $S=\{1,\dots,n\}$ by $a \sim b$ if $c_{abk} \ne 0$ or $c_{bak} \ne 0$ for some $k$. Then the equivalence classes $S_a=\{x:x\sim a\}$ define all possible common subspaces of the form  $C = \mathrm{span} \{\ket{i}:i\in S_x\}.$
\end{proposition}

To identify $C$, prepare two copies of $\psi_{ABR}$, construct a swapped version $\psi'$ by exchanging $C^\perp$, and run a swap test. Let $C$ be the union of subspaces that pass this test. Then, by~\cref{thm3}, this gives the infimum for $u[C]$.

\subsubsection{Decomposed states}

For the lower bound, we classify the structure of entanglement. Consider a decomposition into EPR and GHZ states~\cite{Vidal2000, Lee2024QUI}. Suppose:
\begin{align}\label{eq:tight-state}
\ket{\psi} = &c_1 \frac{1}{\sqrt{2}}(\ket{000} + \ket{101}) + c_2 \frac{1}{\sqrt{2}} (\ket{212} + \ket{223}) \nonumber \\ &+ c_3 \frac{1}{\sqrt{2}} (\ket{334} + \ket{444}) + c_4 \ket{555}.
\end{align}
Then we have:

\begin{proposition}[Reversible entanglement decomposition of tensor powers of tripartite states]\label{prop5}
For $\ket{\psi}^{\otimes n}$, there exists a reversible map $\Lambda_n$ such that
\begin{align}
\ket{\psi}^{\otimes n} &\approx \ket{\mathrm{EPR}}_{AR}^{\otimes \lfloor r_1 n \rfloor} \otimes \ket{\mathrm{EPR}}_{BR}^{ \otimes \lfloor r_2 n\rfloor} \nonumber \\
&\otimes \ket{\mathrm{EPR}}_{AB}^{ \otimes \lfloor r_3 n \rfloor} \otimes \ket{\mathrm{GHZ}}_{ABR}^{ \otimes \lfloor r_4 n \rfloor },
\end{align}
where $r_1 = c_1^2,~ r_2 = c_2^2,~ r_3 = c_3^2,~ r_4 = -\sum_{i=1}^4 c_i^2 \log c_i^2$.
\end{proposition}

It is known that not every tripartite state admits such decomposition~\cite{Acin2003}, and there is no general method to construct it. Thus, we restrict estimation to cases where such decompositions are explicitly given.

\section{Numerical simulations}

%%%1. 기본적인 예측%%%
%%%2. Bound를 표현할 수 있는 figure 넣기%%%

\begin{figure}[t!]
    \centering
    \includegraphics[width=\linewidth]{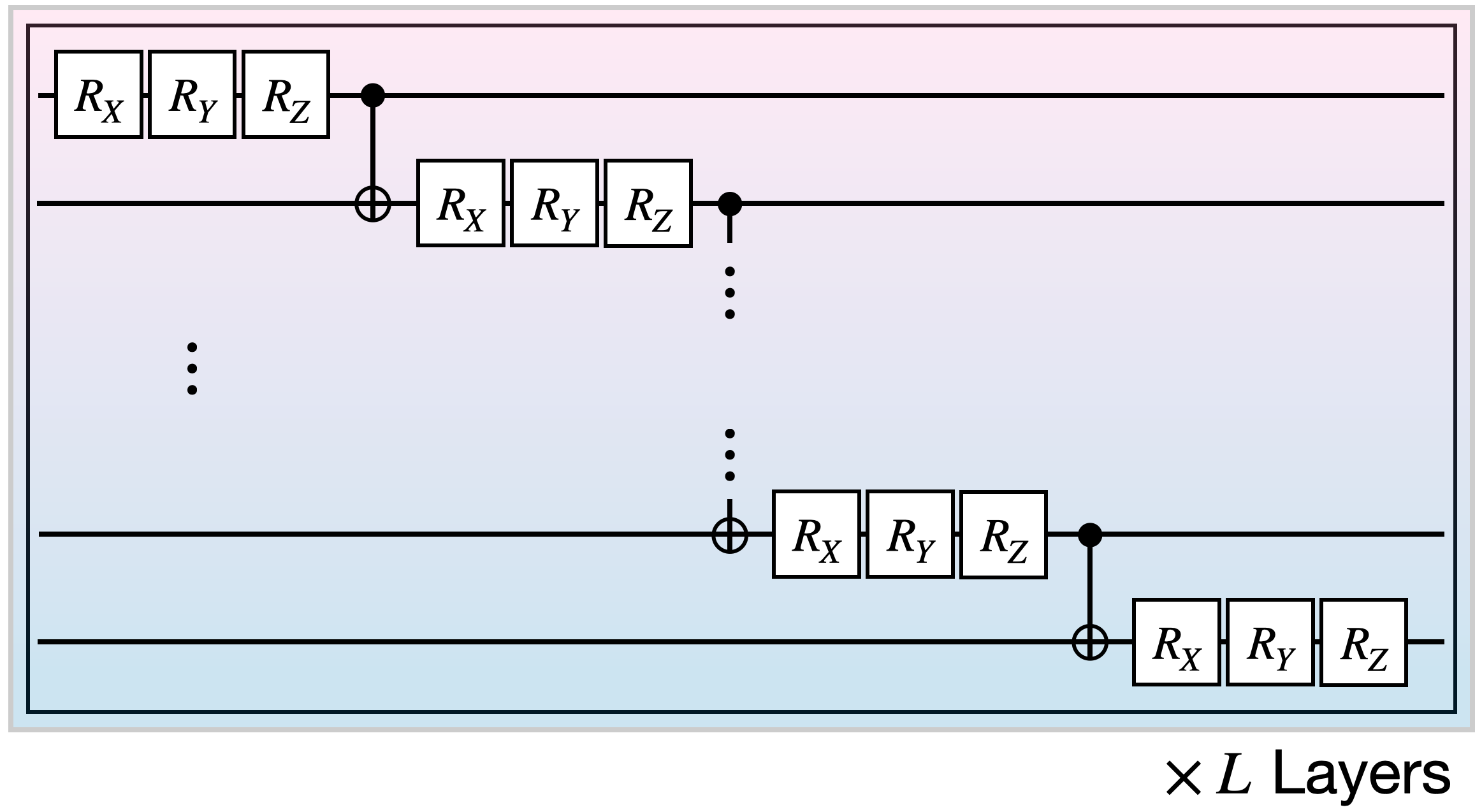}
    \caption{ 
    \textbf{Ansatz structure in numerical simulations.} Shown is the layered ansatz architecture utilized to construct the parameterized unitary $U(\bm{\theta})$ in the numerical experiments. The same structure is uniformly repeated over $L$ layers.}
    \label{fig_ansatz}
\end{figure}

To verify whether the proposed method accurately estimates the bounds, we conducted numerical simulations.
First, we implemented a quantum machine learning algorithm to estimate the von Neumann entropy of a given state. In this setup, the parameterized unitary $U(\bm{\theta})$ was constructed using rotation gates along each axis, together with CNOT gates. The specific structure of the ansatz is shown in~\cref{fig_ansatz}.

Subsequently, we describe the estimation procedures for each of the bounds.  {To verify the performance and stability of our proposed method, we conducted extensive numerical simulations. Each simulation reported in this section was performed for 10 independent runs. For each run, the variational parameters of the quantum circuit were initialized with different random values.}

 {The figures presented in this section depict the aggregated results of these multiple runs. The solid line represents the mean of the entropy estimates at each iteration, while the shaded area corresponds to the standard deviation. This shaded region thus represents the statistical variability of our algorithm's convergence behavior, originating from the different random initial starting points in the optimization landscape.}

\begin{figure}[t!]
\centering
  \begin{subfigure}{\linewidth}
    \centering
    \includegraphics[width=0.7\linewidth]{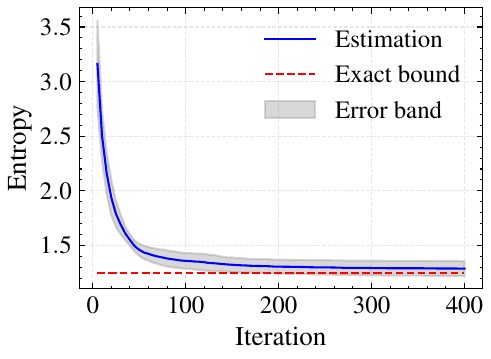}
    \captionsetup{justification=centering}
    \caption{Estimation result for the 4-qubit case.}
  \end{subfigure}

  \vspace{1em}

  \begin{subfigure}{\linewidth}
    \centering
    \includegraphics[width=0.7\linewidth]{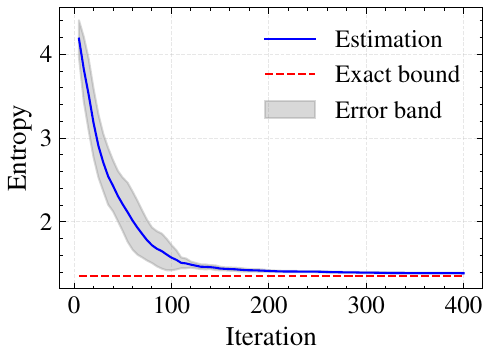}
    \captionsetup{justification=centering}
    \caption{Estimation result for the 6-qubit case.}
  \end{subfigure}

  \vspace{1em}

  \begin{subfigure}{\linewidth}
    \centering
    \includegraphics[width=0.7\linewidth]{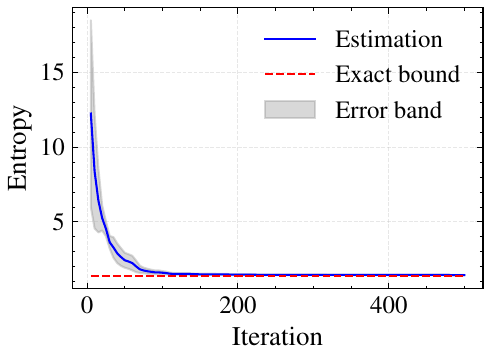}
    \captionsetup{justification=centering}
    \caption{Estimation result for the 8-qubit case.}
  \end{subfigure}
  
    \caption{  \textbf{Convergence of the loose upper bound $S(AB)$.} Shown are the convergence profiles of the upper bound $S(AB)$ for quantum states consisting of 4, 6, and 8 qubits.}
  \label{fig1}
\end{figure}

\begin{figure}[t!]
\centering
  \begin{subfigure}{\linewidth}
    \centering
    \includegraphics[width=0.7\linewidth]{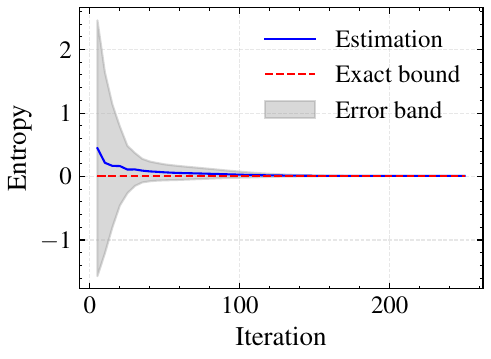}
    \captionsetup{justification=centering}
    \caption{Estimation result for the 4-qubit case.}
  \end{subfigure}

  \vspace{1em}

  \begin{subfigure}{\linewidth}
    \centering
    \includegraphics[width=0.7\linewidth]{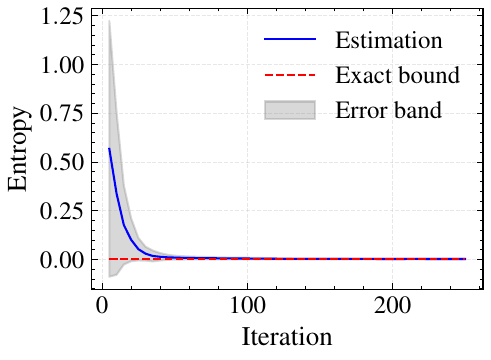}
    \captionsetup{justification=centering}
    \caption{Estimation result for the 6-qubit case.}
  \end{subfigure}
  
  \vspace{1em}

  \begin{subfigure}{\linewidth}
    \centering
    \includegraphics[width=0.7\linewidth]{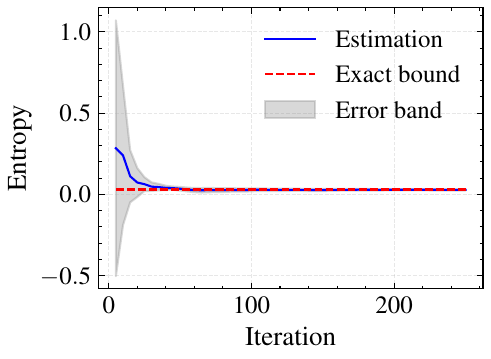}
    \captionsetup{justification=centering}
    \caption{Estimation result of for the 8-qubit case.}
  \end{subfigure}
  
  \caption{  \textbf{Convergence of the loose lower bound $|S(B) - S(A)|$.} Shown are the convergence profiles of the quantity $|S(B) - S(A)|$ as a loose lower bound, evaluated on 4-, 6-, and 8-qubit quantum states.}
  
  \label{fig2}
\end{figure}

\subsection{Estimation of the loose bounds}

For the loose upper bound $S(AB)$ and the loose lower bound $|S(B) - S(A)|$, we directly estimated each von Neumann entropy of the given $\rho_{AB}$.
To simulate this, we first generated random quantum states and partitioned the system equally, assigning half to $A$ and the other half to $B$, thereby constructing the target state for entropy estimation. Then, using the entropy estimation algorithm described in the previous section, we estimated the corresponding entropies. This simulation was conducted for both 4, 6, and 8 qubit systems.

The simulation results for the upper bound $S(AB)$ are shown in~\cref{fig1}. The estimation of $S(AB)$ stabilized after approximately 100 optimization steps for the 4-qubit system and around 200 steps for the 6-qubit and 8-qubit system.

For the lower bound $|S(B)-S(A)|$, since only half of the system is used to estimate $S(A)$ and $S(B)$, the actual subsystems involved correspond to 2-qubit, 3-qubit, and 4-qubit reduced states, respectively. The results are presented in~\cref{fig2}.

The estimation of $|S(B) - S(A)|$ stabilized after roughly 50 optimization steps. Theoretically, since the error in the quantum Donsker–Varadhan representation-based method is proportional to the rank of the density matrix, larger systems require more optimization steps to achieve stable estimates.

\begin{figure}[t!]
\centering
  \begin{subfigure}{\linewidth}
    \centering
    \includegraphics[width=0.7\linewidth]{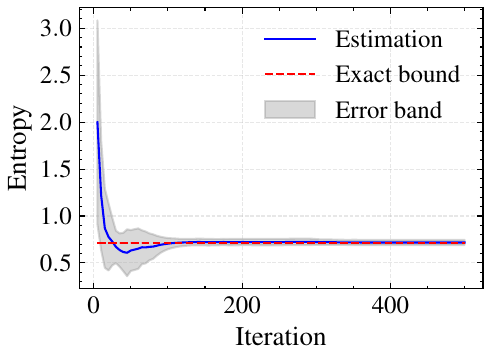}
    \captionsetup{justification=centering}
    \caption{Estimation result for the 4-qubit case.}
  \end{subfigure}

  \vspace{1em}
  
  \begin{subfigure}{\linewidth}
    \centering
    \includegraphics[width=0.7\linewidth]{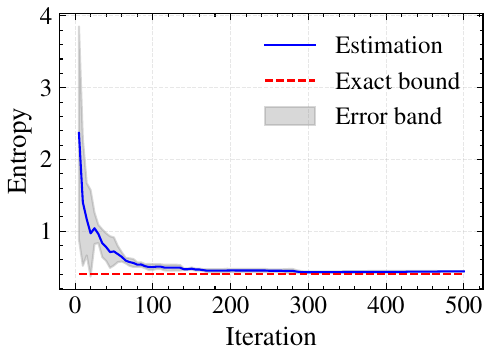}
    \captionsetup{justification=centering}
    \caption{Estimation result for the 8-qubit case.}
  \end{subfigure}
  
  \caption{   \textbf{Convergence of the tight upper bound $u[C]$.} Shown are the convergence profiles of the tight upper bound $u[C]$ evaluated on 4- and 8-qubit quantum states.}
  
  \label{fig3}
\end{figure}

\subsection{Estimation of the tight upper bound}
\begin{figure*}[t!]
\centering
    \includegraphics[width=0.8\linewidth]{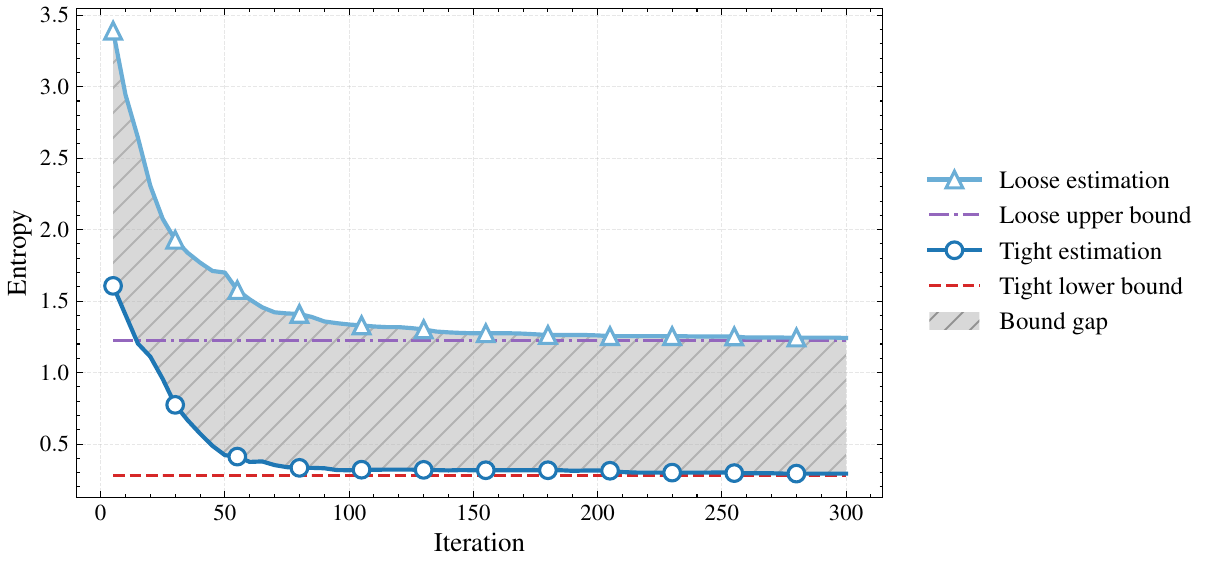}
  
  \caption{  \textbf{Effect of the bound gap on convergence.} Shown is the difference in convergence behavior when a gap exists between the loose and tight bounds.}
  
  \label{fig4}
\end{figure*}

\begin{figure}[t!]
\centering
  \begin{subfigure}{\linewidth}
    \centering
    \includegraphics[width=0.7\linewidth]{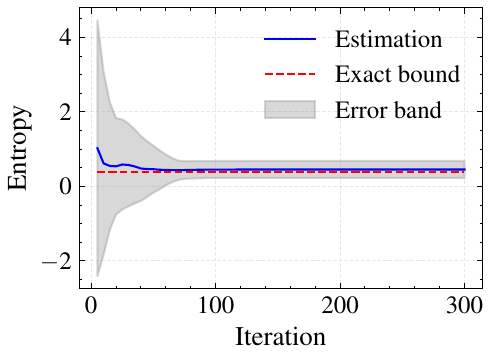}
  \end{subfigure}
  
  \caption{  \textbf{Convergence of the tight lower bound $l[\Lambda]$.} Shown is the convergence of the tight lower bound $l[\Lambda]$ for a given quantum state.}
  
  \label{fig5}
\end{figure}

To estimate the tight upper bound $u[C] = S(R|A)_{\psi_s}$, we first identify the common subspace $C$ of systems $A$ and $B$ from the given state $\psi_{ABR}$, then construct the stretched state $\psi_s$, and perform entropy estimation on the full system $AA'BB'R$.  
In practice, when the system size is small, it is often the case that no nontrivial common subspace exists. Therefore, in this simulation, we fixed the common subspace in advance and proceeded with the estimation.

Specifically, let the dimensions of systems $A$ and $B$ be $2^n$, and fix an integer $1 \leq k \leq 2^n$. Denote the computational basis of $A$ and $B$ by $\{\ket{1}, \dots, \ket{2^n}\}$, and define the common subspace $C$ as $C = \mathrm{span}\{\ket{1}, \dots, \ket{k}\}$. In other words, for the tripartite system $ABR$, we ensured that the components supported on $C \otimes C \otimes R$ were symmetric under the exchange of systems $A$ and $B$, while the components supported on $C^{\perp} \otimes C^{\perp} \otimes R$ were randomly generated.

Simulations were conducted for total system sizes of 4 and 8 qubits for $ABR$. For the 4-qubit system, we set $k = 1$; for the 8-qubit system, we set $k = 2$. The results are shown in~\cref{fig3}.

The estimation of $u[C]$ stabilized after approximately 200 optimization steps. Since the ranks of the reduced states on $AR$ and $A$, which are used in the entropy estimation process, depend on the value of $k$, the number of steps required for convergence increases with the size of the common subspace.

Finally,~\cref{fig4} compares the estimation results for the loose and tight upper bounds of the quantum uncommon information of the given state. The difference between the tight and loose bounds leads to distinct convergence behaviors in their corresponding estimated values.

\subsection{Estimation of the tight lower bound}

For the tight lower bound  $l[\Lambda] = r_1 S(A_1)_{\psi_1} + r_2 S(B_1)_{\psi_2} + r_4 ( S(B_3 R_3)_{\psi_4} - S(A_3 R_3)_{\psi_4})$, the simulation was conducted as described earlier, assuming that the state is given by~\cref{eq:tight-state}.

According to~\cref{prop5}, the state can be decomposed into EPR and GHZ states, which allows us to estimate the value of the lower bound. Specifically, the coefficients $c_1, \dots, c_4$ are randomly chosen such that $\ket{\psi}$ remains a pure state. Based on these coefficients, the values of $r_1, \dots, r_4$ can be computed. Since $\psi_1, \psi_2, \psi_3$ are EPR states and $\psi_4$ is a GHZ state, we estimate the corresponding entropies to obtain the value of $l[\Lambda]$. \cref{fig5} presents the simulation results. The estimation of $l[\Lambda]$ stabilized after roughly 50 optimization steps.

\section{Concluding remarks}

In this paper, we proposed a quantum machine learning approach to estimate the bounds of the quantum uncommon information, which represents the minimum amount of entanglement required to exchange the given state $\rho_{AB}$ between subsystems $A$ and $B$. Since quantum Donsker–Varadhan representation allows the von Neumann entropy of a state to be expressed as the infimum of a specific function, we designed a machine learning algorithm that utilizes this as a cost function to estimate entropy. As the bounds of quantum uncommon information are expressed in terms of von Neumann entropy, the same algorithm can be used to estimate their values. Notably, the number of copies required for training with quantum Donsker–Varadhan representation scales as $\mathrm{poly}(r)$ with respect to the rank $r$, providing an advantage in estimating bounds that involve entropies of subsystems of a given state. In addition, the method proposed in this paper for identifying the common subspace of a given state can be extended to other estimation techniques. 

Our estimation method requires different types of information from Alice and Bob depending on the bound to be estimated. Fundamentally, the use of quantum Donsker–Varadhan representation requires knowledge of the rank of each state. In scenarios where the ranks are not known, one may employ techniques such as quantum rank estimation~\cite{Ryan2019VQD}.

To determine the common subspace, Alice and Bob must additionally know the marginal states they individually possess and how the entire system can be decomposed. In cases where the marginal states are not known, one can employ a variational quantum circuit to search for unitaries $U$ and $W$. However, this approach still requires access to information about the global system.

Moreover, a decomposed state requires prior knowledge of whether the given state can be expressed in a specific canonical form. Therefore, depending on the context in which quantum uncommon information is applied, one must estimate different bounds accordingly.

Quantum uncommon information can be broadly applied in scenarios where the parties are required to exchange their quantum states completely, including entanglement, and thus provides a natural measure of communication cost in such contexts. These situations frequently arise during entanglement distribution protocols in quantum networks~\cite{AppQN1997, AppQN2016, App1999DQC, App2004DQC, App2013DQC, App2016DQC, App2017QKD, App2020QKD, App2023QKD, App1999QSS, App2008QSS}.

In many of these cases, Alice and Bob have prior knowledge about their respective states, particularly when the state preparation process is known. Furthermore, in these cases, the states often consist of EPR or GHZ states. Under these conditions, the requirements for identifying the common subspace and the decomposed state are often satisfied, and the estimation method for quantum uncommon information can be utilized effectively.

To allow for more general application, certain aspects need to be improved. First, if the process of identifying the common subspace can be refined to work without prior knowledge of the structure of the original state, then it may be possible to determine the common subspace based solely on information from the partial systems held by Alice and Bob. Furthermore, our approach remains applicable even when different forms of decomposed states are considered.

 {Although this work assumes an ideal quantum system, the presence of noise in realistic quantum devices could affect the performance of the proposed methodology. For instance, errors in state preparation or gate operations within the variational circuit could degrade the accuracy of the loss function calculation, potentially introducing a bias to the entropy estimate or hindering convergence.}

 {A crucial direction for future research is therefore to integrate quantum error mitigation techniques into our algorithm to test its robustness in noisy environments. Such an extension would significantly enhance the practical applicability of the proposed method.}

%%%추후 연구 과제?%%%

\section*{Data availability statement}

The data and software that support the findings of this study can be found in the following repository: \url{https://github.com/donghwa722/QUINE}

\section*{Acknowledgements}

J.L. acknowledges helpful discussions with Ju-Young Ryu.
This work was supported by the National Research Foundation of Korea (NRF) through a grant funded by the Ministry of Science and ICT (Grant No. RS-2025-00515537). This work was also supported by the Institute for Information \& Communications Technology Promotion (IITP) grant funded by the Korean government (MSIP) (Grant Nos. RS-2019-II190003 and RS-2025-02304540), the National Research Council of Science \& Technology (NST) (Grant No. GTL25011-401), and the Korea Institute of Science and Technology Information (KISTI) (Grant No. P25026).  
I.K.S. acknowledges support by Quantum Computing based on Quantum Advantage challenge research through the National Research Foundation of Korea (NRF) funded by the Korean government (MSIT) (Grant No. RS-2023-00256221).

%%%%%%%%%%%%%%%%%%%%%%%%%%%%%%%%%%%%%%%%%%%%%
\bibliography{reference}

\end{document}